\documentclass[a4paper,11pt]{article}
\usepackage{pos}

\title{Z production in pPb collisions at LHCb}

\author*[a,1]{Hengne Li}

\affiliation[a]{Guangdong Provincial Key Laboratory of Nuclear Science, 
Institute of Quantum Matter, South China Normal University, 
Guangzhou 510006, China}

\note{On behalf of the LHCb collaboration.}

\emailAdd{hengne.li@m.scnu.edu.cn}

\abstract{

This article presents results  
of the Z boson production in the proton-lead collisions 
at $\sqrt{s_{\rm NN}} = 5.02$~TeV and $\sqrt{s_{\rm NN}}= 8.16$~TeV 
collected in 2013 and 2016, respectively, 
by the LHCb detector at the LHC in forward and backward rapidity. 
The great precision of the 2016 data are found compatible with 
nPDFs theoretical predictions within large theoretical uncertainties, 
and can be useful to constraint new predictions.
}

\FullConference{%
  HardProbes2020\\
  1-6 June 2020\\
  Austin, Texas}

\begin{document}
\maketitle

The properties of W/Z bosons have been extensively 
studied at electron-positron and hadron colliders. 
The production cross-sections of W/Z bosons at hadron colliders
can be well described by perturbative Quantum Chromodynamics (pQCD) 
 at next-to-next-to-leading order (NNLO), 
 and the radiative corrections and the input electroweak parameters 
are also precisely known.
The measurements of their production cross-sections 
in proton-proton (pp) collisions can be used
to constrain the initial conditions such as 
the Parton Distribution Functions 
(PDFs) of the proton~\cite{Rojo:2015acz, Butterworth:2015oua}.

In the same respect, 
the production of the W/Z bosons can also precisely probe
of the nuclear PDFs, especially the that of heavy quarks and gluon, 
which are currently less precisely constrained~\cite{Kusina:2016fxy}.
In proton-ion and ion-ion collisions, 
the PDFs of nucleons confined 
in nuclei are found to be different with respect to those of free nucleons,
which are called 
nuclear PDFs (nPDFs)~\cite{Eskola:2016oht, deFlorian:2003qf, Hirai:2007sx, AtashbarTehrani:2012xh, Khanpour:2016pph}. 
The differences between nPDFs and PDFs are often referred as nuclear modifications,
which are understood as  
a reflection of the various initial state nuclear matter effects on the free nucleons.
These effects include the nuclear shadowing~\cite{Glauber:1955qq}
appearing as a suppression 
for Bjorken-$x$ ($x$ in the following, the fraction of a nucleon momentum carried by a parton) below 0.05 , 
the anti-shadowing effect~\cite{Ashman:1988bf, Brodsky:1989qz} raising the PDFs for $x$ around 0.1 , 
the EMC effect~\cite{Aubert:1983xm} suppressing the PDFs around $0.3<x<0.7$ , 
and the fermi motion effect modifies the region for $x$ around 1. 
The nPDFs are crucial for the studies of the Quark Gluon Plasma (QGP) in the ion-ion collisions,
in order to disentangle cold and hot nuclear matter effects.

Moreover, 
since the W/Z bosons and their leptonic decay products
do not participate strong interactions, 
they inherit perfectively the initial conditions 
without being modified by the hadronic medium in the intermediate and final states. 
Therefore, they can be used to better differentiate between properties of 
the initial- and final-state effects,
and the proton-ion collisions provide 
an ideal environment to study the initial-state nuclear matter effects, 
hence to constrain the nPDFs.


In this article, we present results  
of the Z boson production in the proton-lead collisions~\cite{Aaij:2014pvu, LHCb:CONF2019003} at the LHC 
using data collected during 2013 and 2016 by the LHCb detector. 
The LHCb detector~\cite{LHCb-DP-2008-001, LHCb-DP-2014-002} 
is a fully instrumented single-arm spectrometer in the forward region 
covering a pseudorapidity acceptance of
$2 < \eta < 5$,  providing a high tracking momentum resolution down to very low 
transverse momentum ($p_{\rm T}$) and precise vertex reconstruction capability. 
The proton-lead datasets with their
recorded integrated luminosities are given in Table~\ref{tab:dataset}.

\begin{table}[htbp]
\begin{center}
\begin{tabular}{c|cccc}
                       & \multicolumn{2}{c}{2013}       &   \multicolumn{2}{c}{2016}      \\  
$\sqrt{s_{\rm NN}}$ &  \multicolumn{2}{c}{5.02\,TeV}  &  \multicolumn{2}{c}{8.16\,TeV}   \\ 
                        \hline
                        & pPb     &  Pbp         &   pPb   &  Pbp    \\

$\mathcal{L}$ & 1.1\,nb$^{-1}$ & 0.5\,nb$^{-1}$ &13.6\,nb$^{-1}$ &20.8\,nb$^{-1}$  \\
\end{tabular}
\end{center}
\vspace*{-0.5cm}
\caption{Summary of the LHCb pPb datasets and the recorded integrated luminosities.}
\label{tab:dataset}
\end{table}

The Z boson production cross-sections in the dimuon decay channel are measured in the 
fiducial volume in both the forward (pPb) and backward (Pbp) collision 
configurations~\cite{Aaij:2014pvu, LHCb:CONF2019003} 
based on the following equation:
$\sigma_{{\rm Z}\to\mu^+\mu^-} = \left[{\rm N_{cand}}\cdot \rho\right]/\left[\mathcal{L}\cdot \epsilon\right]$,
where $\sigma_{{\rm Z}\to\mu^+\mu^-}$ is the production cross-section to be measured,
${\rm N_{cand}}$ is the number of $ {\rm Z}\to\mu^+\mu^-$ candidates passing 
signal selection, 
$\rho$ is the signal purity of the selected Z candidates,
$\mathcal{L}$ is the integrated luminosity, 
and $\epsilon$ is the total efficiency
including trigger, reconstruction and selection efficiencies. 
The fiducial volume is defined as $60<m_{\mu^+\mu^-}<120$\,GeV,
$2.0<\eta_{\mu^{\pm}}<4.5$, and $p_{\rm T}^{\mu^{\pm}}>20$\,GeV.
The purity is measured using data-driven methods, and the efficiencies are estimated 
using Monte-Carlo (MC) samples together with tag-and-probe data driven corrections.

\begin{figure}[h]
  \begin{center}
    \includegraphics[width=0.4\linewidth]{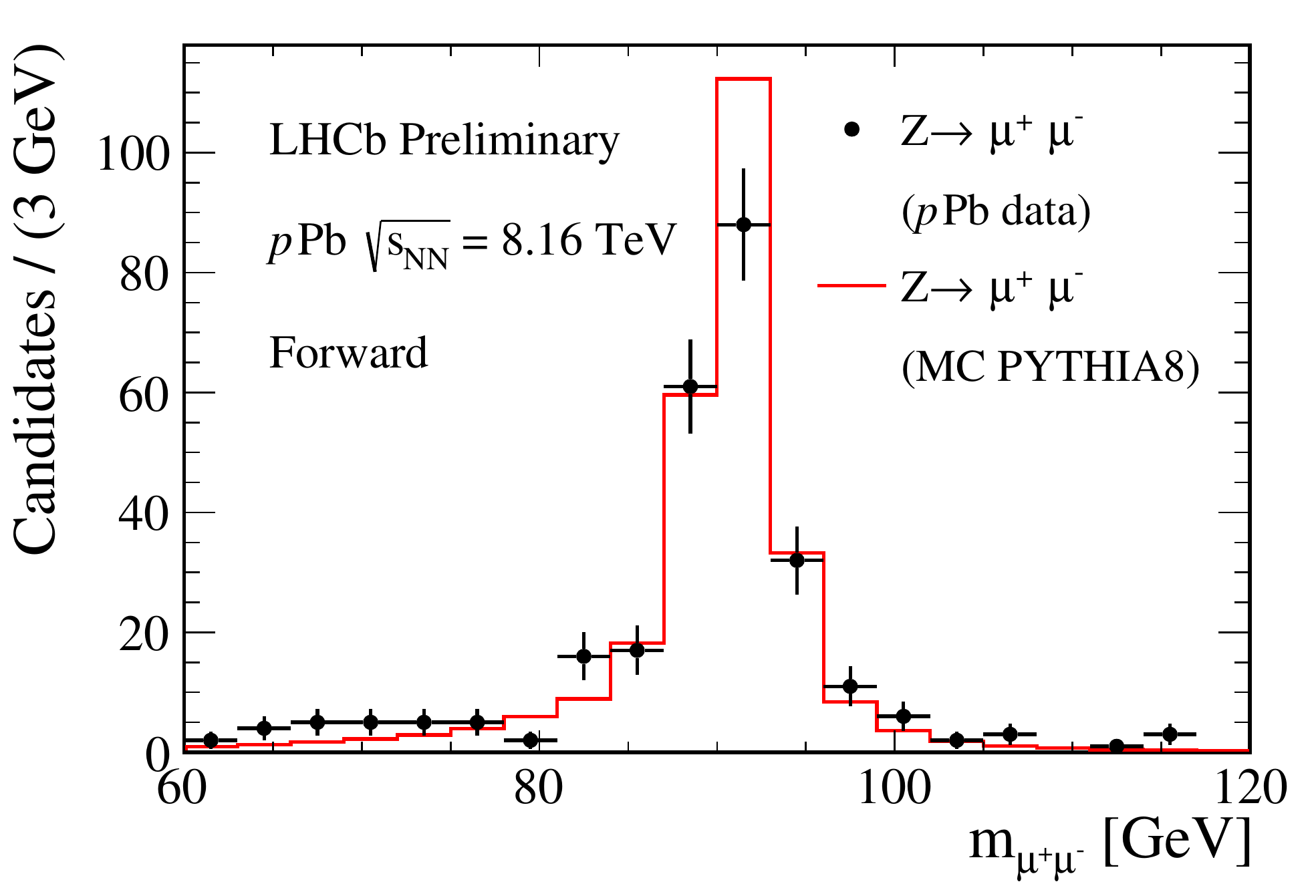}
    \put(-130,56){(a)}
    \includegraphics[width=0.4\linewidth]{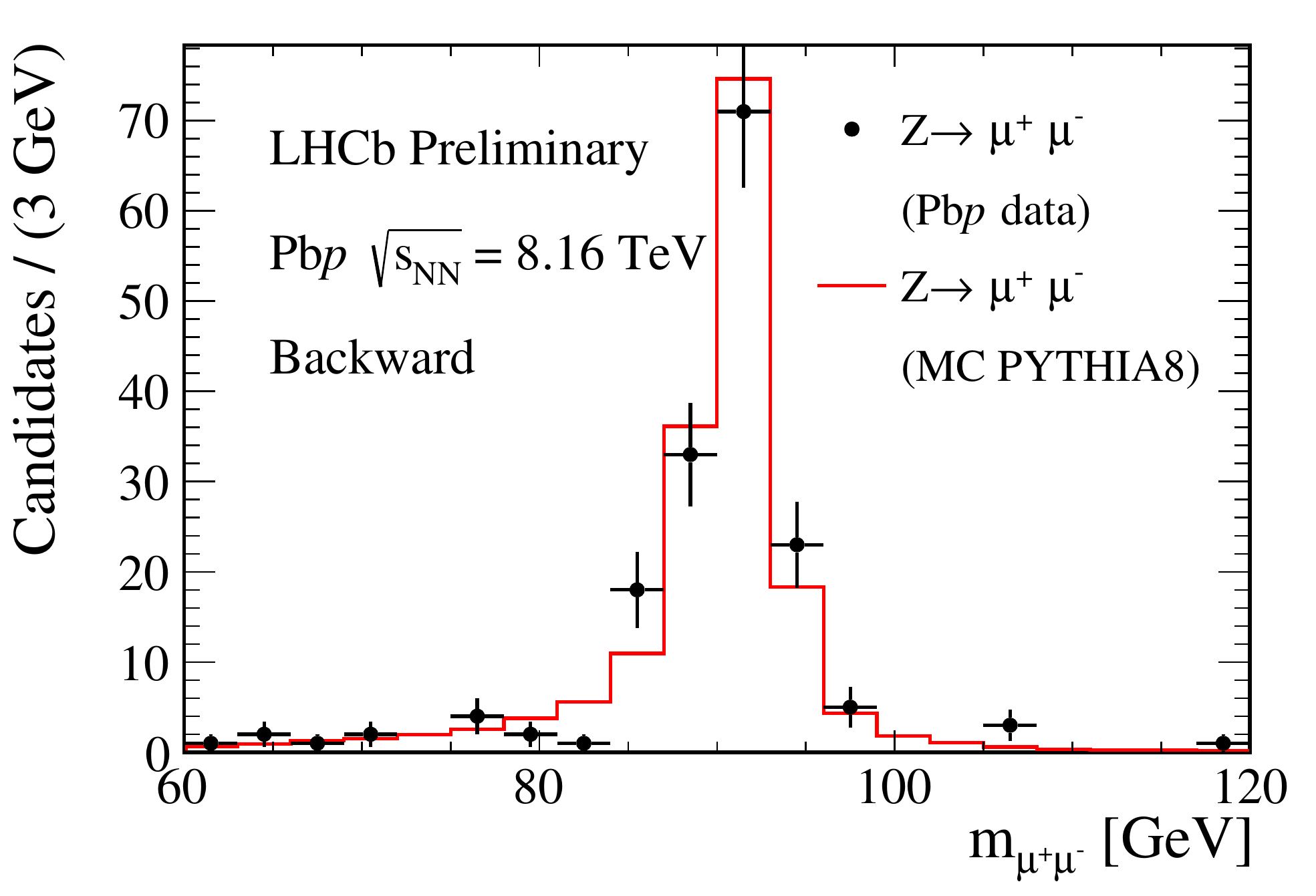}
    \put(-130,56){(b)}
    \vspace*{-0.5cm}
  \end{center}
  \caption{(color online) The dimuon invariant mass distributions after the offline selection
for pPb (a) and Pbp (b) configurations,
using datasets taken 
in 2016 at $\sqrt{s_{\rm NN}} = 8.16$\,TeV.
The red line shows the distributions from 
simulation generated using PYTHIA 8~\cite{Sjostrand:2007gs} with CTEQ6L1~\cite{Stump:2003yu} PDF set, 
normalised to the number of observed candidates. }
  \label{fig:candidates}
\end{figure}

The invariant mass distributions of selected signal candidates
are shown in Fig.~\ref{fig:candidates} for datasets 
taken in 2016 at $\sqrt{s_{\rm NN}} = 8.16$\,TeV.

\begin{figure}[h]
\begin{center}
 \includegraphics[width=0.4\linewidth]{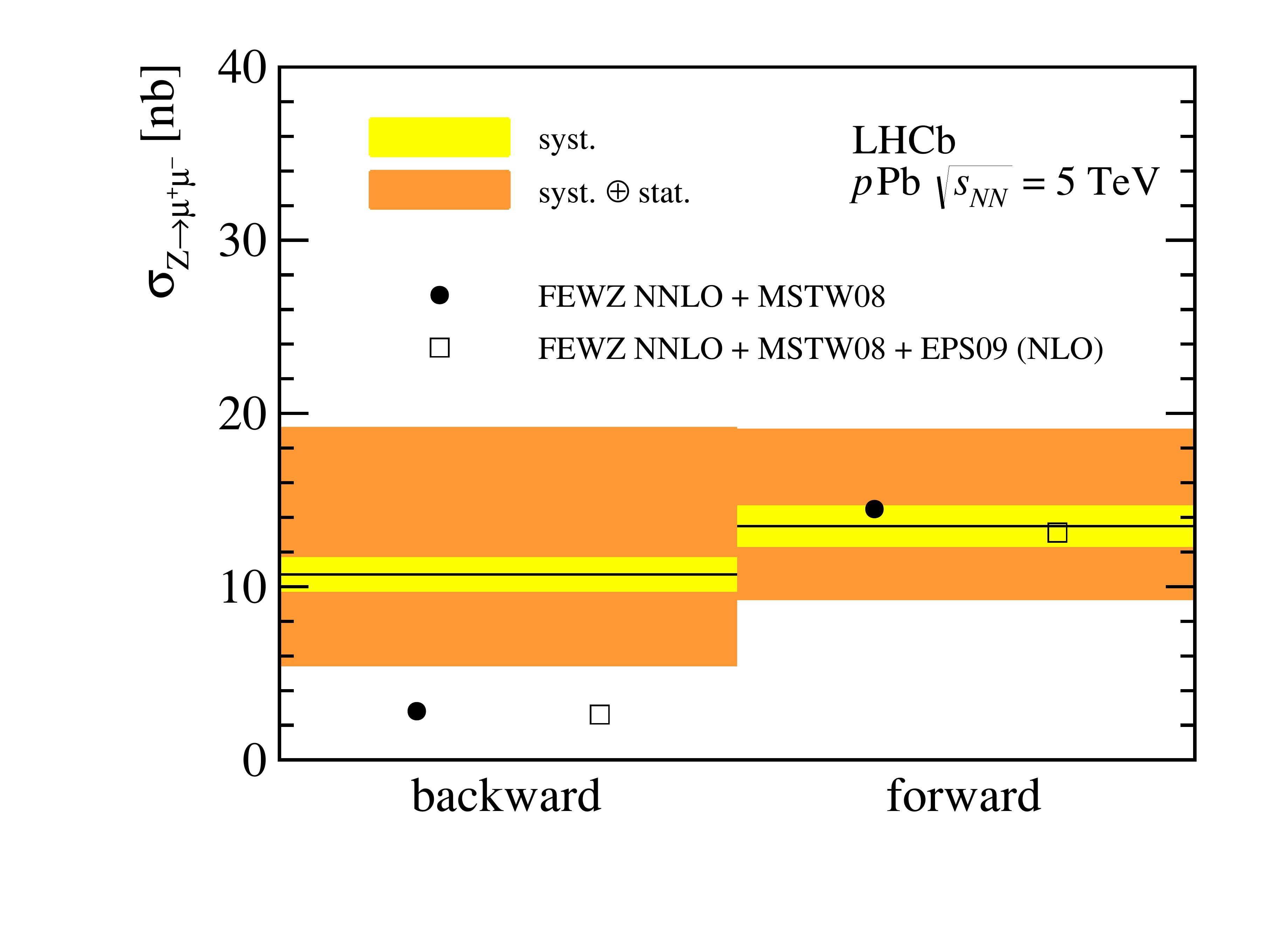}
     \put(-85,7){(a)}
\includegraphics[width=0.42\linewidth]{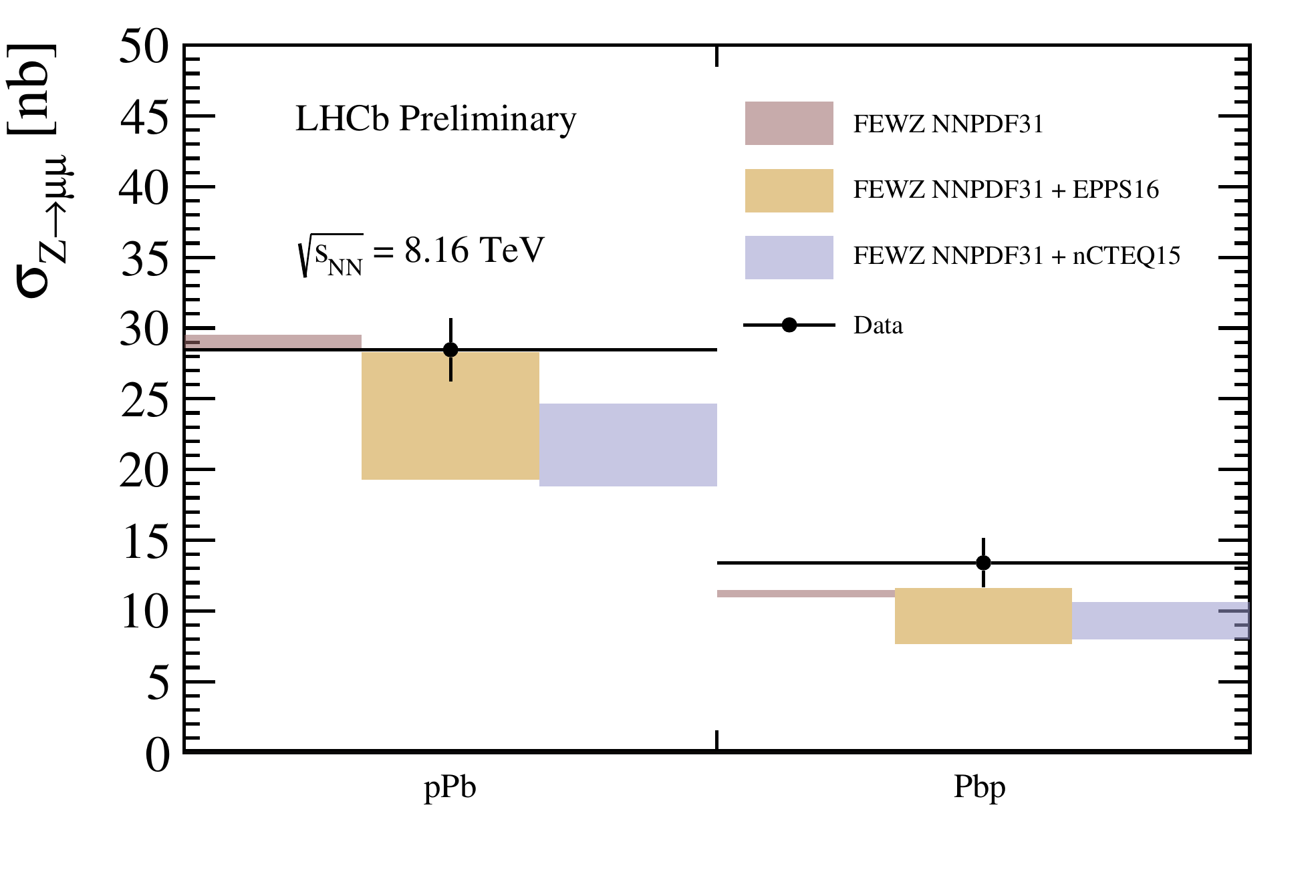}
     \put(-90,7){(b)}
\vspace*{-0.8cm}
\end{center}
\caption{(color online) Measured fiducial cross-sections of Z production
and theoretical calculations using various PDF sets with or without nuclear modification.
Figure (a) is for dataset taken in 2013 at $\sqrt{s_{\rm NN}} = 5.02$\,TeV  
and figure (b) is for dataset taken in 2016 at $\sqrt{s_{\rm NN}} = 8.16$\,TeV.}
  \label{fig:result_cross_section}
\end{figure}

\begin{figure}[h]
\begin{center}
  \includegraphics[width=0.50\linewidth]{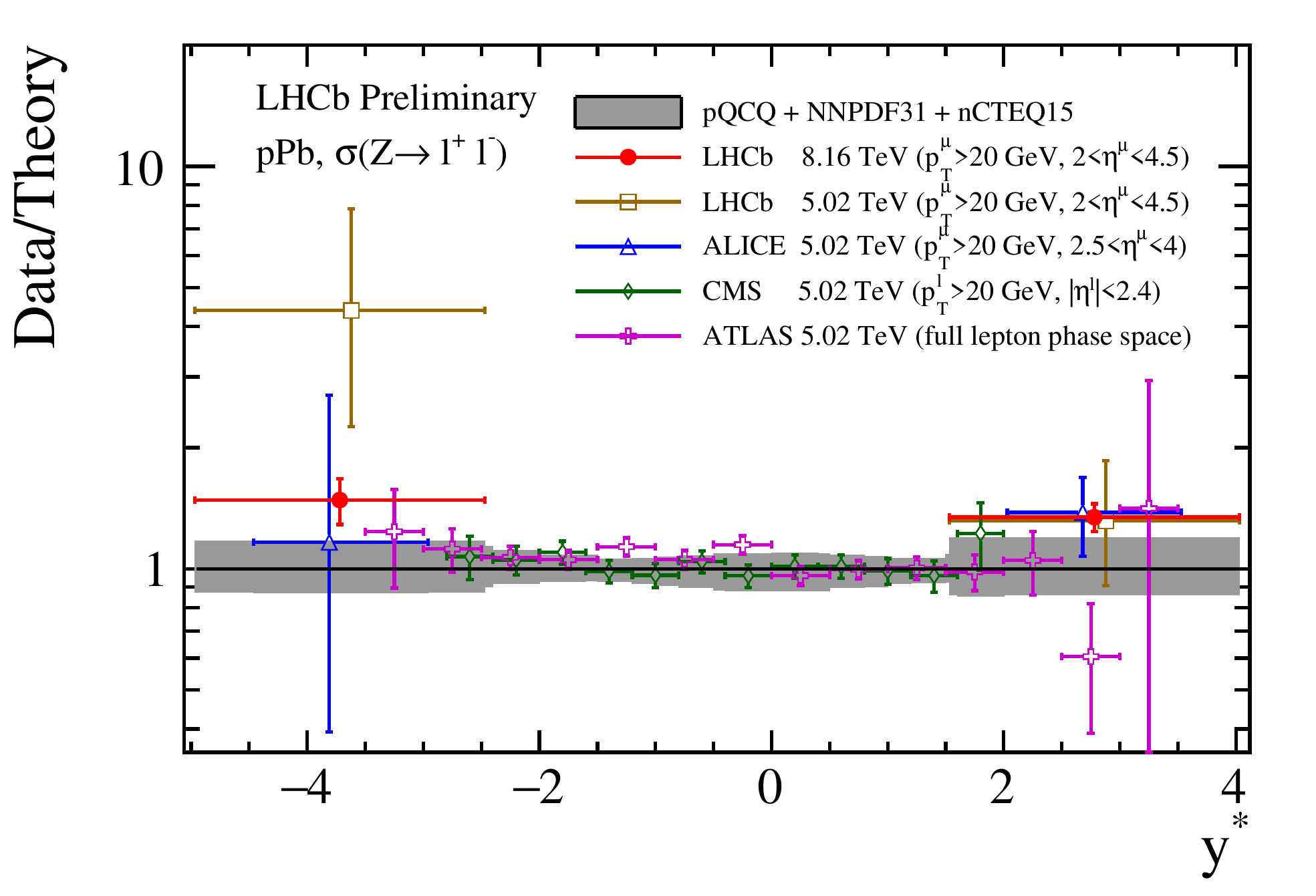}
\vspace*{-0.8cm}
\end{center}
\caption{(color online) Comparison of LHCb 8.16\,TeV results with previous 5.02\,TeV results from ATLAS, 
CMS, ALICE, and LHCb. 
The uncertainties on the data over theory ratios include only the experimental statistical and systematic uncertainties; 
the PDF uncertainties are shown separately on the line at one by the grey band. 
The central values of the LHCb and ALICE results at 5.02\,TeV are shifted to left and right by 
0.1 units in rapidity, respectively, for better visibility.}
  \label{fig:result_cross_section_compare}
\end{figure}

The resulting fiducial cross-sections for centre-of-mass energies 
at $\sqrt{s_{\rm NN}} =$ 5.02\,TeV and 8.16\,TeV~\cite{Aaij:2014pvu, LHCb:CONF2019003} 
are shown in Fig.~\ref{fig:result_cross_section} (a) and (b), respectively.
The results are compared with theoretical calculations using
FEWZ~\cite{Gavin:2010az,Li:2012wna} with MSTW08~\cite{Martin:2009iq} free nucleon PDF 
together with EPS09 (NLO)~\cite{Eskola:2009uj} nPDF for dataset at 5.02\,TeV,
and with NNPDF 3.1~\cite{Ball:2017nwa} free nucleon PDF together EPPS16 (NLO)~\cite{Eskola:2016oht} 
 and nCTEQ15 (NLO)~\cite{Kusina:2016fxy, Kovarik:2015cma} nPDFs
for dataset at 8.16\,TeV. 
The measured cross-section central value is higher than the nPDFs predicted values 
but in agreement with the theoretical prediction statistically.  
The results are also compared with previous 5.02\,TeV results from 
LHCb~\cite{Aaij:2014pvu}, ATLAS~\cite{Aad:2015gta},
CMS~\cite{Khachatryan:2015pzs}, and ALICE~\cite{Alice:2016wka}, 
as shown in Fig.~\ref{fig:result_cross_section_compare}.
The new LHCb 8.16\,TeV results are compatible with 
previous 5.02\,TeV results, but with about 20 times higher statistics.
The great precision of these measurements at forward and backward 
rapidities are able to provide strong constraint, specially at backward rapidity.

The ratio of the Z boson production cross-sections 
for forward and backward configurations ($R_\mathrm{FB}$), 
is particularly sensitive to cold nuclear effects. 
$R_\mathrm{FB}$ is measured in the common 
rapidity region ($2.5<|y^*|<4.0$) in the
centre-of-mass frame of the produced Z boson 
using 2016 dataset~\cite{LHCb:CONF2019003}  as
    $R_\mathrm{FB}^{2.5<|y^*|<4.0}= 1.28\pm0.14({\rm stat})\pm0.14({\rm syst})\pm0.05({\rm lumi})$,
which is compatible with theoretical calculations using FEWZ with the following nPDFs:
 $R_\mathrm{FB,NNPDF3.1+EPPS16}^{2.5<|y^*|<4.0} = 1.45 \pm 0.10{\rm (theo.)} \pm 0.01 {\rm (num.)} \pm 0.27 {\rm (nPDF)}$, and 
 $R_\mathrm{FB,NNPDF3.1+nCTEQ15}^{2.5<|y^*|<4.0} = 1.44 \pm 0.10{\rm (theo.)} \pm 0.01 {\rm (num.)} \pm 0.20 {\rm (nPDF)}$,
where, the uncertainty ``num.'' is from the numerical precision.

In summary, 
LHCb provides an excellent opportunity to probe the cold nuclear matter effects 
in the very forward region using Z boson production.
Results of pPb collisions at 5.02\,TeV and 8.16\,TeV results are presented,
which are compatible with theoretical predictions involving nPDFs, where the
8.16\,TeV results give the highest precision in the forward region at LHC,
thus can be useful in constraining the current nPDFs.

\begin{footnotesize}

\acknowledgments
We acknowledge the support from Science and Technology Program of Guangzhou (No. 2019050001).

%

\bibliographystyle{JHEP}
\bibliography{main}
\end{footnotesize}

\end{document}